\documentstyle[12pt]{article}
\begin{document}
\begin{center}
{\bf{SPECIAL PROPERTIES OF A SUBMANIFOLD OF $su(3)$ }}

\vspace{1.0cm}

R.Parthasarathy{\footnote{e-mail address:
sarathy@imsc.ernet.in}} \\
The Institute of Mathematical Sciences \\
Chennai 600 113, India. \\
\end{center}

\vspace{1.0cm}

{\noindent{\it{Abstract}}}

The finite gauge transformation, in the submanifold of $su(3)$
introduced in hep-th/9902027, is found to be {\it{linear}} 
in the gauge
functions ${\omega}^a$'s. In this manner, a parameterization
of $SU(3)$ in terms of a single real octet vector ${\omega}^a$
is achieved.  
Under this finite transformation,
those gauge fields satisfying $D^{ab}_{\mu}{\omega}^b\ =\ 0$,
remain unchanged. A geometric meaning to the submanifold in
terms of subgroup of rotations in $E8$ is presented.  

\vspace{1.0cm}

In a recent paper [1], dealing with the infrared region of QCD
and confinement, $SU(3)$ gauge field configurations
appropriate to the infrared region of QCD have been proposed.
These configurations, motivated by the works of 'tHooft [2]
and Mandelstam [3] on the crucial roles of monopole
configurations in confinement, are defined by
\begin{eqnarray}
D^{ab}_{\mu}{\omega}^b \ =\ {\partial}_{\mu}{\omega}^a +
f^{acb}\ {A}^c_{\mu}\ {\omega}^b &=& 0,
\end{eqnarray}
where, the eight components ${\omega}^a$ of an eight
dimensional vector  
are chosen to satisfy
\begin{eqnarray}
{\omega}^a{\omega}^a & = & 1, \nonumber \\
d^{abc}\ {\omega}^b\ {\omega}^c &=& \frac{1}{\sqrt{3}}\ 
 {\omega}^a.
\end{eqnarray}
Now, the manifold relevant to the infrared region of QCD is
{\t{not}} the complete $su(3)$ manifold spanned by arbitrary
${\omega}^a$'s, 
{\it{but}} a submanifold spanned by those ${\omega}^a$'s
satisfying (2) and the relevant gauge field configuration
$A^a_{\mu}$ are {\it{determined}} by solving (1). In [1], a
solution $A^a_{\mu}$ satisfying (1) and (2) has been found to
be
\begin{eqnarray}
A^a_{\mu} &=& C_{\mu} {\omega}^a -
\frac{4}{3}f^{abc}{\omega}^b\ {\partial}_{\mu}{\omega}^c,
\end{eqnarray}
where $C_{\mu}$ is an arbitrary vector field. Further it has
been shown that such configurations when used in the usual QCD
action contains magnetic monopole configurations necessary for
confinement and confinement has been realized as in the London
theory of Meissner effect, along the lines suggested by Nambu
[4]. It has been stated in [1] that in view of (1), a mass
term $m^2\ A^a_{\mu}\ A^a_{\mu}$ for the gauge field
$A^a_{\mu}$ (satisfying (1)) can be added without violating
gauge symmetry, as a gauge variation of this term $2m^2(\delta
A^a_{\mu})\ A^a_{\mu}\ =\ 2m^2\ D^{ab}_{\mu}{\omega}^b\
A^a_{\mu}$ identically vanishes in view of (1), as long as we
are working in the submanifold defined by (2).

\vspace{0.5cm}

The purpose of this short note is to show this result of
infinitesimal variation holdsgood for finite gauge
transformations as well. The finite gauge transformations here
become {\it{linear}} in ${\omega}^a$'s. The second relation in
(2) upon using the first relation becomes
$d_{abc}{\omega}^a{\omega}^b{\omega}^c\ =\
\frac{1}{\sqrt{3}}$. It is known [5,6] that
${\omega}^a{\omega}^a$ and
$d_{abc}{\omega}^a{\omega}^b{\omega}^c$ are the {\it{only}} two $SU(3)$
invariants. The submanifold is thus one in which these
invariants are taken as constants unity and
$\frac{1}{\sqrt{3}}$ respectively.  

\vspace{0.5cm}

Consider a gauge transformation generated by
\begin{eqnarray}
U &=& exp\left(i{\omega}^a \frac{{\lambda}_a}{2}\right),
\end{eqnarray}
for ${\omega}$'s satisfying (2) and ${\lambda}_a$'s are the
Gell-Mann $SU(3)$ matrices. We shall make use of the basic
relation for the $\lambda $- matrices, viz., 
\begin{eqnarray}
{\lambda}_a {\lambda}_b &=& if^{abc}{\lambda}_c +
\frac{2}{3}{\delta}^{ab} + d^{abc}{\lambda}_c,
\end{eqnarray}
where $f^{abc}$ and $d^{abc}$ are the usual anti-symmetric and
symmetric $SU(3)$ tensors respectively. Expanding the
exponential (4), we have
\begin{eqnarray}
U&=& 1 + i{\omega}^a\frac{{\lambda}_a}{2} - \frac{1}{2!}\
\frac{1}{4}\ ({\omega}^a{\lambda}_a)^2 \nonumber \\
&-& \frac{i}{3!}\ \frac{1}{8}\ ({\omega}^a{\lambda}_a)^3
+ \frac{1}{4!}\ \frac{1}{16}\ ({\omega}^a{\lambda}_a)^4 +
\cdots . \nonumber  
\end{eqnarray}
Using (5), it follows
\begin{eqnarray}
({\omega}^a{\lambda}_a)^2 &=& \frac{2}{3}\ +\
\frac{1}{\sqrt{3}}\ {\omega}^a{\lambda}_a, \nonumber \\
({\omega}^a{\lambda}_a)^3 &=& \frac{2}{3\sqrt{3}} + {\omega}^a
{\lambda}_a, \nonumber \\
({\omega}^a{\lambda}_a)^4 &=& \frac{2}{3} \ +\ \frac{5}{3
\sqrt{3}}\ {\omega}^a{\lambda}_a, \nonumber \\
({\omega}^a{\lambda}_a)^5 &=& \frac{10}{9\sqrt{3}} \ +\ \frac
{11}{9}\ {\omega}^a{\lambda}_a, \nonumber 
\end{eqnarray}
and so on. Substituting these in $U$, we have,
\begin{eqnarray}
U&=& (1-\frac{1}{2!}\ \frac{1}{6}-\frac{i}{3!}\
\frac{1}{12\sqrt{3}}+\frac{1}{4!}\
\frac{5}{144\sqrt{3}}+\cdots ) \nonumber \\
&+&(\frac{i}{2}-\frac{1}{2!}\ \frac{1}{4\sqrt{3}}-\frac{i}
{3!}\ \frac{1}{8}+\frac{1}{4!}\ \frac{5}{48\sqrt{3}}+\cdots )
\ {\omega}^a{\lambda}_a \nonumber \\
&=& \alpha \ +\ \beta \ {\omega}^a{\lambda}_a,
\end{eqnarray}
where $\alpha$ and $\beta$ are (finite) complex constants.
$U^{-1}$ can be easily found to be ${\alpha}^*\ +\ {\beta}^*
{\omega}^a{\lambda}_a$ (where $^*$ stands for complex
conjugation) and so $U{U}^{-1}\ =\ I$, gives the conditions,
\begin{eqnarray}
\alpha {\alpha}^*\ +\ \frac{2}{3}\ \beta {\beta}^* &=& 1,
\nonumber \\
\alpha\ {\beta}^*\ +\ \beta\ {\alpha}^*\ +\ \frac{1}{\sqrt{3}}
\ \beta\ {\beta}^* &=& 0.
\end{eqnarray}
Thus, the finite gauge transformation $U$ becomes
{\it{linear}} in ${\omega}$. This is a special feature of the
submanifold defined by (2). {\it{Thus a parameterization of
$SU(3)$ in terms of a single real octet vector ${\omega}^a$ is
achieved by (6).}}  
It has been shown by Macfarlane,
Sudbery and Weisz [5] that {\it{any}} special unitary matrix $U$
written as $exp(iA)$ with $A$ hermitian and $A\ =\
{\omega}_a{\lambda}_a$, can be written as $U\ =\ u_0 \ +\
iu_a{\lambda}_a$, where $u_a\ =\ a{\omega}_a + b
d_{abc}{\omega}^b{\omega}^c$, with $u_0,a,b$ as functions of
the two $SU(3)$ invariants, ${\omega}^a{\omega}^a$ and
$d_{abc}{\omega}^a{\omega}^b{\omega}^c$. In our choice, these
invariants are taken as constants $1$ and $\frac{1}{\sqrt{3}}$
and this gives (6) for $U$.  
We will relate this $U$ to the
rotation matrix of a subgroup of rotations in eight
dimensional Euclidean space $E8$ at the end.  

\vspace{0.5cm}

Under a finite gauge transformation, the gauge field
$A_{\mu}\ =\ A^a_{\mu}\frac{{\lambda}_a}{2}$ transforms as
\begin{eqnarray}
A_{\mu}\ \rightarrow \ A^U_{\mu}&=& \frac{1}{i}\
({\partial}_{\mu}U)U^{-1} + U\ A_{\mu}\ U^{-1}.
\end{eqnarray}
To check the notation and the factors, we find the
infinitesimal version of (8) is
\begin{eqnarray}
{\delta}A^a_{\mu} &=& D^{ac}_{\mu}{\omega}^c,
\end{eqnarray}
as it should be. In view (1), $\delta A^a_{\mu}\ =\ 0$ and we
want to show that for {\it{finite}} gauge transformations (6)
with (7),
\begin{eqnarray}
A^{U}_{\mu} &=& A_{\mu},
\end{eqnarray}
for those $A^a_{\mu}$'s satisfying (1). 

\vspace{0.5cm}

We shall make use of the standard relations among $f$ and $d$
tensors [5], viz.,
\begin{eqnarray}
f^{i\ell m}d^{mjk} + f^{j\ell m}d^{imk} + f^{klm}d^{ijm} &=&
0,
\end{eqnarray}
and 
\begin{eqnarray}
f^{ijm}f^{k\ell m}&=&\frac{2}{3}({\delta}^{ik}{\delta}^{j\ell}
- {\delta}^{i\ell}{\delta}^{jk})+d^{ikm}d^{j\ell m}-d^{jkm}
d^{i\ell m}.
\end{eqnarray}

\vspace{0.5cm}

Starting from (6), we have
\begin{eqnarray}
({\partial}_{\mu}U) U^{-1}&=&\beta
({\partial}_{\mu}{\omega}^a){\lambda}_a({\alpha}^*+{\beta}^*
{\omega}^b{\lambda}_b) \nonumber \\
&=&\beta {\alpha}^*({\partial}_{\mu}{\omega}^a){\lambda}_a+
i\beta{\beta}^*({\partial}_{\mu}{\omega}^a){\omega}^bf^{abc}
{\lambda}_c \nonumber \\  
&+&\beta{\beta}^*({\partial}_{\mu}{\omega}^a)
{\omega}^bd^{abc}{\lambda}_c,
\end{eqnarray}
where $({\partial}_{\mu}{\omega}^a){\omega}^a\ =\ 0$,
following from the first relation in (2) is used. Also, the
second relation in (2) gives
\begin{eqnarray}
d^{abc}({\partial}_{\mu}{\omega}^b){\omega}^c &=&
\frac{1}{2\sqrt{3}}\ ({\partial}_{\mu}{\omega}^a),
\end{eqnarray}
and so
\begin{eqnarray}
({\partial}_{\mu}U)U^{-1}&=&
(\beta{\alpha}^*+\frac{1}{2\sqrt{3}}\beta{\beta}^*)
({\partial}_{\mu}
{\omega}^a){\lambda}_a+i\beta{\beta}^*({\partial}_{\mu}
{\omega}^a){\omega}^bf^{abc}{\lambda}_c. 
\end{eqnarray}
We are interested in the transformation of the $A^a_{\mu}$'s
satisfying (1) and so using (1) in (15), we have
\begin{eqnarray}
({\partial}_{\mu}U)U^{-1}&=&(\beta{\alpha}^*+\frac{1}
{2\sqrt{3}}
\beta{\beta}^*)f^{abc}{\omega}^bA^c_{\mu}{\lambda}_a
\nonumber \\
&+& i\beta{\beta}^* (f^{apq}{\omega}^pA^q_{\mu}){\omega}^b
f^{abc}{\lambda}_c.
\end{eqnarray}
Using (13) for simplifying $f^{apq}f^{abc}$, (16) becomes
\begin{eqnarray}
({\partial}_{\mu}U)U^{-1}&=&(\beta{\alpha}^*+\frac{1}{2\sqrt
{3}}\beta{\beta}^*)f^{abc}{\omega}^bA^c_{\mu}{\lambda}_a
\nonumber \\
&+& i\beta{\beta}^*\{ \frac{2}{3}A^a_{\mu}{\lambda}_a -
\frac{2}{3}{\omega}^b{\omega}^cA^b_{\mu}{\lambda}_c + \frac{1}
{\sqrt{3}}{\omega}^a\ d^{qca}A^q_{\mu}{\lambda}_c \nonumber
\\
&-& d^{qba}d^{pca}{\omega}^p{\omega}^bA^q_{\mu}{\lambda}_c\},
\end{eqnarray}
where the relations in (2) have been used.      

\vspace{0.5cm}

Now, we consider the evaluation of $UA_{\mu}U^{-1}$ using (6).
\begin{eqnarray}
UA_{\mu}U^{-1}&=& (\alpha + \beta {\omega}^a{\lambda}_a)\
A^b_{\mu}\frac{{\lambda}_b}{2}\ ({\alpha}^* + {\beta}^*
{\omega}^c{\lambda}_c), \nonumber \\
&=& \frac{1}{2}(\alpha + \beta {\omega}^a{\lambda}_a)\
\{{\alpha}^* {\lambda}_b + i{\beta}^* f^{bcd}\ {\omega}^c 
{\lambda}_d + \frac{2}{3}{\beta}^* {\omega}^b \nonumber \\
&+& {\beta}^* d^{bcd}\ {\omega}^c {\lambda}_d\} A^b_{\mu},
\end{eqnarray}
using (5). Expanding further and using (5), we have
\begin{eqnarray}
UA_{\mu}U^{-1}&=& \frac{1}{2}\{\alpha {\alpha}^*{\lambda}_b +
i\alpha {\beta}^* f^{bcd} {\omega}^c{\lambda}_d + \frac{2}{3}
\alpha {\beta}^* {\omega}^b \nonumber \\
&+&\alpha{\beta}^*d^{bcd}{\omega}^c{\lambda}_d+\beta{\alpha}^*
{\omega}^a(if^{abc}{\lambda}_c+\frac{2}{3}{\delta}_{ab}+
d^{abc}{\lambda}_c) \nonumber \\
&+& i\beta{\beta}^*f^{bcd}{\omega}^c{\omega}^a(if^{ade}
{\lambda}_e+\frac{2}{3}{\delta}_{ad}+d^{ade}{\lambda}_e)
\nonumber \\
&+& \frac{2}{3}\beta{\beta}^*\ {\omega}^a{\omega}^b{\lambda}_a
\nonumber \\
&+&\beta{\beta}^*d^{bcd}{\omega}^a{\omega}^c(if^{ade}{\lambda}
_e+\frac{2}{3}{\delta}_{ad}+d^{ade}{\lambda}_e)\}A^b_{\mu}.
\end{eqnarray} 
In here there are fourteen terms. The sixth term and the third
term are like-terms; the seventh and the fourth are
like-terms, and the fifth and the second are like-terms. The
nineth term vanishes identically. In the thirteenth term, the
second relation in (2) is used. Then we find,
\begin{eqnarray}
UA_{\mu}U^{-1}&=&\frac{1}{2}\{\alpha{\alpha}^*{\lambda}_b+i(
\alpha{\beta}^*-\beta{\alpha}^*)f^{bcd}{\omega}^c{\lambda}_d+
\frac{2}{3}(\alpha{\beta}^*+\beta{\alpha}^*){\omega}^b
\nonumber \\
&+& (\alpha{\beta}^* + \beta{\alpha}^*)
d^{bcd}{\omega}^c{\lambda}_d \nonumber \\
&+&i\beta{\beta}^*(if^{ade}f^{bcd}{\omega}^a{\omega}^c
{\lambda}_e+f^{bcd}d^{ade}{\omega}^c{\omega}^a{\lambda}_e)
\nonumber \\
&+& \frac{2}{3}\beta{\beta}^*{\omega}^a{\omega}^b{\lambda}_a
\nonumber \\
&+&\beta{\beta}^*(if^{ade}d^{bcd}{\omega}^a{\omega}^c{\lambda}
_e+\frac{2}{3\sqrt{3}}{\omega}^b+d^{bcd}d^{ade}{\omega}^a
{\omega}^c{\lambda}_e)\}A^b_{\mu}.
\end{eqnarray} 
Now, the coefficient of ${\omega}^b$ terms in (20) is
$\frac{2}{3}\ (\alpha{\beta}^*\ +\ \beta{\alpha}^*\ +\
\frac{1}{\sqrt{3}}\beta{\beta}^*)$, which is zero due to the
second relation in (7). Using the relation (12) to rewrite the
eighth term $f^{ade}d^{bcd}$ as $-f^{bad}d^{edc}\ -\
f^{cad}d^{ebd}$, the term involving $f^{cad}$ vanishes due to
the symmetry of ${\omega}^a{\omega}^c$ in $a$ and $c$. Then
the remaining of the eighth term cancells with the sixth term
in (20). In the fifth term involving $f^{ade}f^{bcd}$, we make
use of (13). Then (20) becomes,
\begin{eqnarray}
UA_{\mu}U^{-1}&=&\frac{1}{2}\{\alpha{\alpha}^*{\lambda}_b
+i(\alpha{\beta}^*-\beta{\alpha}^*)f^{bcd}{\omega}^c
{\lambda}_d+(\alpha{\beta}^*+\beta{\alpha}^*)d^{bcd}
{\omega}^c{\lambda}_d \nonumber \\
&-&\beta{\beta}^*(\frac{2}{3}{\lambda}_b-\frac{2}{3}{\omega}
^b{\omega}^c{\lambda}_c+\frac{d^{ebd}}{\sqrt{3}}{\omega}^d
{\lambda}_e-d^{abd}d^{ecd}{\omega}^a{\omega}^c{\lambda}_e)
\nonumber \\
&+&\frac{2}{3}\beta{\beta}^*{\omega}^a{\omega}^b{\lambda}_a
+\beta{\beta}^*d^{bcd}d^{ade}{\omega}^a{\omega}^c{\lambda}
_e\}A^b_{\mu}.
\end{eqnarray} 
Here, the two terms involving the two $d$-tensors are the same
and so,
\begin{eqnarray}
UA_{\mu}U^{-1}&=&\frac{1}{2}\{\alpha{\alpha}^*{\lambda}_b
+i(\alpha{\beta}^*-\beta{\alpha}^*)f^{bcd}{\omega}^c
{\lambda}_d+(\alpha{\beta}^*+\beta{\alpha}^*)d^{bcd}
{\omega}^c{\lambda}_d \nonumber \\
&-&\frac{2}{3}\beta{\beta}^*{\lambda}_b+\frac{4}{3}\beta
{\beta}^*{\omega}^a{\omega}^b{\lambda}_a-\beta{\beta}^*
\frac{1}{\sqrt{3}}d^{ebd}{\omega}^d{\lambda}_e \nonumber \\
&+& 2\beta{\beta}^* d^{abd}d^{ecd}{\omega}^a{\omega}^c
{\lambda}_e\}\ A^b_{\mu}.
\end{eqnarray}
From (17) and (22), we have,
\begin{eqnarray}
\frac{1}{i}({\partial}_{\mu}U)U^{-1}+UA_{\mu}U^{-1}&=&-i(\beta
{\alpha}^*+\frac{1}{2\sqrt{3}}\beta{\beta}^*)f^{abc}{\omega}
^b{\lambda}_aA^c_{\mu} \nonumber \\
&+&\beta{\beta}^*\{\frac{2}{3}A^a_{\mu}{\lambda}_a-\frac{2}
{3}{\omega}^b{\omega}^cA^b_{\mu}{\lambda}_c \nonumber \\
&+&\frac{1}{\sqrt{3}}d^{qca}{\omega}^aA^q_{\mu}{\lambda}_c
-d^{qba}d^{pca}{\omega}^p{\omega}^bA^q_{\mu}{\lambda}_c\}
\nonumber \\
&+&\frac{1}{2}\{\alpha{\alpha}^*{\lambda}_b+i(\alpha{\beta}^*
-\beta{\alpha}^*)f^{bcd}{\omega}^c{\lambda}_d \nonumber \\
&+&(\alpha{\beta}^*+\beta{\alpha}^*)d^{bcd}{\omega}^c
{\lambda}_d-\frac{2}{3}\beta{\beta}^*{\lambda}_b+\frac{4}{3}
\beta{\beta}^*{\omega}^a{\omega}^b{\lambda}_a \nonumber \\
&-&\frac{\beta{\beta}^*}{\sqrt{3}}d^{ebd}{\omega}^d{\lambda}
_e+2\beta{\beta}^*d^{abd}d^{ecd}{\omega}^a{\omega}^c{\lambda}
_e\}A^b_{\mu}.
\end{eqnarray}  
 
\vspace{0.5cm}

Now, consider the terms involving two $d$-tensors. They are,
\begin{eqnarray}
(-d^{qba}d^{pca}{\omega}^p{\omega}^bA^q_{\mu}{\lambda}_c\ +\
d^{abd}d^{ecd}{\omega}^a{\omega}^cA^b_{\mu}{\lambda}_c).
\nonumber 
\end{eqnarray}
By changing the summation indices in the second term above,
viz., $b\rightarrow q, \ a\rightarrow b,\ c\rightarrow p,\
e\rightarrow c$ in that order, the two terms cancell each
other. Then, the terms involving one $d$-tensor are,
\begin{eqnarray}
\beta{\beta}^*d^{qca}{\omega}^aA^q_{\mu}{\lambda}_c+\frac{1}
{2}(\alpha{\beta}^*+\beta{\alpha}^*)d^{bcd}{\omega}^c
{\lambda}_dA^b_{\mu}-\frac{\beta{\beta}^*}{2\sqrt{3}}d^{ebd}
{\omega}^d{\lambda}_eA^b_{\mu}. \nonumber  
\end{eqnarray}
By rearranging the indices, this becomes,
\begin{eqnarray}
\frac{1}{2}(\alpha{\beta}^*+\beta{\alpha}^*+\frac{1}{\sqrt{3}}
\beta{\beta}^*)d^{qca}{\omega}^aA^q_{\mu}{\lambda}_c,
\nonumber 
\end{eqnarray}
which vanishes due to the second relation in (7). Now,
consider the terms involving one $f$-tensor. They are,
\begin{eqnarray}
-i(\beta{\alpha}^*+\frac{\beta{\beta}^*}{2\sqrt{3}})f^{abc}
{\omega}^bA^c_{\mu}{\lambda}_a+\frac{i}{2}(\alpha{\beta}^*
-\beta{\alpha}^*)f^{bcd}{\omega}^c{\lambda}_dA^b_{\mu}.
\nonumber 
\end{eqnarray}
In the second term, make $b\leftrightarrow c,\ d\rightarrow a$
so that the above expression becomes,
\begin{eqnarray}
-\frac{i}{2}(\alpha{\beta}^*+\beta{\alpha}^*+\frac{1}{\sqrt{3}
}\beta{\beta}^*)f^{abc}{\omega}^bA^c_{\mu}{\lambda}_a,
\nonumber 
\end{eqnarray}
which vanishes due to the second relation in (7). Finally, the
terms involving two ${\omega}$'s in (23) cancell each other,
leaving
\begin{eqnarray}
\frac{1}{i}({\partial}_{\mu}U)U^{-1}+UA_{\mu}U^{-1}&=&
\frac{1}{2} (\alpha{\alpha}^*+\frac{2}{3}\beta{\beta}^*)
A^a_{\mu}{\lambda}_a\ =\ \frac{1}{2}A^a_{\mu}{\lambda}_a,
\nonumber \\
&=& A_{\mu},
\end{eqnarray}
using the first relation in (7).

\vspace{0.5cm}

Thus when $U\ =\ exp\{i{\omega}^a\frac{{\lambda}_a}{2}\}$
{\it{finite}} transformation is considered, $U$ becomes
$\alpha +\beta {\omega}^a{\lambda}_a$,  {\it{linear}} in
${\omega}^a$ in view of (2), with $\alpha$ and $\beta$
satisfying (7). This finite transformation leaves $A_{\mu}$
invariant, for those $A^a_{\mu}$'s satisfying (1). Therefore,
within the submanifold defined by (2), the gauge fields
satisfying (1) can have a mass term, even for finite
transformations.

\vspace{0.5cm}

It will be useful to consider the converse of this result. In
order to this, we ask the question: What will be the gauge
field configurations that remain unchanged under finite gauge
transformations? Since under a finite gauge transformation
$U$, $A_{\mu}$ transforms as,
\begin{eqnarray}
A_{\mu}\ \rightarrow \ A^U_{\mu} &=&
\frac{1}{i}({\partial}_{\mu}U)U^{-1} + UA_{\mu}U^{-1},
\nonumber
\end{eqnarray}
the answer is given by
\begin{eqnarray}
A_{\mu} &=&
\frac{1}{i}({\partial}_{\mu}U)U^{-1}+UA_{\mu}U^{-1}.  \nonumber
\end{eqnarray}
By right multiplying the above expression by $U$, we have
\begin{eqnarray}
\frac{1}{i}{\partial}_{\mu}U + [U,A_{\mu}] &=& 0,
\end{eqnarray}
whose infinitesimal version is (1). This means that for those
$A_{\mu}$'s and $U$ satisfying (25), the mass term remain
invariant. In order to give explicit expressions for
$A^a_{\mu}$ and $U$, the transformation $U$ is chosen by those
${\omega}$'s satisfying (2). Then (6) gives the required $U$.
The choice of ${\omega}$'s satisfying (2) and $A^a_{\mu}$ in
(3) determined by (1),  
produce magnetic
monopole configurations in the QCD action [1].  

\vspace{0.5cm}

It is known that the tensor indices taking eight values in $d$
and $f$ are tensor indices associated with the adjoint group
$SU(3)/Z(3)$ of $SU(3)$. The tensors $d$ and $f$ associated
with $SU(3)/Z(3)$, are cartesian tensors in eight real
dimensions [6]. Given a single octet vector $\{ {\omega}^a\
\}$ and the tensors $d$ and $f$, it is known that {\it{at
most}} two linearly independent octets can be formed. They are
${\omega}^a$ itself and $d_{abc}{\omega}^b{\omega}^c$. Then
{\it{at most}} two independent $SU(3)$ invariants can be
formed, which are taken as, 
\begin{eqnarray}
{\omega}^a{\omega}^a, \nonumber \\
d_{abc}{\omega}^a{\omega}^b{\omega}^c.
\end{eqnarray} 
A geometric meaning can be given to the invariants. If we
associate a $3\times 3$ matrix $A$ with ${\omega}^a$ as $A\
=\ {\omega}^a\ {\lambda}_a$, where ${\lambda}^a$'s are the
Gell-Mann matrices, then $Tr(A^2)$ and $det(A)$ give the above
two invariants. In our choice made in (2), we have taken the
two invariants as constants. 

\vspace{0.5cm}

A study of the relationship of the adjoint group $SU(3)/Z(3)$
to the subgroup of rotation group $R8$ which leaves invariant
the length ${\omega}^a{\omega}^a$ of the real eight component
vector ${\omega}^a$ and the cubic invariant
$d_{abc}{\omega}^a{\omega}^b{\omega}^c$ has been made by
Macfarlane [6]. Here we will give the main results for our
choice (2). The eight components ${\omega}^a$ can be taken to
describe a point of $E8$. Rotations in $E8$ are real linear
transformations
\begin{eqnarray}
{\omega}^a\ \rightarrow \ {\omega}^{'a} &=& R_{ab}{\omega}^b.
\end{eqnarray}
Invariance of the length leads to $RR^T\ =\ I$. We take
$det(R)\ =\ 1$. To relate $SU(3)$ to a subgroup of $R8$, the
group of rotations in $E8$, associate with each point in $E8$,
a $3\times 3$ traceless hermitian matrix A
\begin{eqnarray}
A &=& {\omega}^a{\lambda}_a,
\end{eqnarray}
an element of the algebra of $SU(3)$. Transformations of $E8$
induced by $U\ \in \ SU(3)$ tansformation
\begin{eqnarray}
A\ \rightarrow\ A' &=& UAU^{-1},
\end{eqnarray}
can be shown to give [6] 
\begin{eqnarray}
R_{ab} &=& \frac{1}{2}\ Tr\{ {\lambda}_a \ U\ {\lambda}_b\
U^{-1}\}.
\end{eqnarray}
For $U$ in (6) and using the relations in (7), we find
explicitly,
\begin{eqnarray}
R_{ab}&=&(1-{\beta}{\beta}^*){\delta}_{ab}+2{\beta}{\beta}^*
{\omega}_a{\omega}_b-\sqrt{3}{\beta}{|beta}^*d_{abc}{\omega}
_c \nonumber \\
&+& if_{abc}{\omega}_c\ (\alpha{\beta}^*-\beta{\alpha}^*).
\end{eqnarray}
It is verified that $R_{ab}R_{ac}\ =\ {\delta}_{bc}$, i.e.,
the transformation is orthogonal. It is seen that
$R_{ab}{\omega}_b\ =\ {\omega}_a$. Thus the rotation leaves
${\omega}^a{\omega}^a$ invariant. It is seen that the
invariance of $d_{abc}{\omega}^a{\omega}^b{\omega}^c$ is
verified by showing $d_{abc}R_{ap}R_{bq}R_{cr}\ =\ d_{pqr}$.

\vspace{0.5cm}

I am extremely thankful to Biswajit Chakraborty, Ramesh
Anishetty and H.Sharatchandra for discussions.  
 
\vspace{0.5cm}

{\noindent{\bf{References.}}}

\begin{enumerate}

\item R.Parthasarathy, hep-th/9902027.

\item G.'tHooft, Nucl.Phys. {\bf B190}, (1981) 455; {\bf
B138}, (1978)1; {\bf B153}, (1979) 141. 

\item S.Mandelstam, Phys.Rev. {\bf D19}, (1978) 2391; in the
{\it{Proceedings of the Monopole Meeting}}, Trieste, Italy,
1981. Edited by N.S.Craige, P.Goddard and W.Nahm, World
Scientific, 1982.

\item Y.Nambu, Phys.Rev. {\bf D10}, (1974) 4262; Phys.Rep.
{\bf C23}, (1975) 250.

\item A.J.Macfarlane, A.Sudbury and P.H.Weisz, Comm.Math.Phys.
{\bf 11}, (1968) 77.

\item A.J.Macfarlane, Comm.Math.Phys. {\bf 11}, (1968) 91. 
\end{enumerate}

\end{document}